\documentclass[amssymb,amsmath,aps,pre,preprint,showpacs]{revtex4} 
\usepackage{graphicx}

\def\func#1{\textrm{#1}}

\begin{document}

\title{Long-Ranged Correlations in Sheared Fluids}
\author{James F. Lutsko}
\affiliation{Center for Nonlinear Phenomena and Complex Systems\\
Universit\'{e} Libre de Bruxelles\\
Campus Plaine, CP 231, 1050 Bruxelles, Belgium}
\author{James W. Dufty}
\affiliation{Deptartment of Physics, University of Florida\\
Gainesville, FL 32611}
\date{\today}

\begin{abstract}
The presence of long-ranged correlations in a fluid undergoing uniform shear
flow is investigated. An exact relation between the density autocorrelation
function and the density-mometum correlation function implies that the
former must decay more rapidly than $1/r$, in contrast to predictions of
simple mode coupling theory. Analytic and numerical evaluation of a
non-perturbative mode-coupling model confirms a crossover from $1/r$
behavior at ''small'' $r$ to a stronger asymptotic power-law decay. The
characteristic length scale is $\ell \approx \sqrt{\lambda _{0}/a}$ where $%
\lambda _{0}$ is the sound damping constant and $a$ is the shear rate.
\end{abstract}

\maketitle

\section{Introduction}

Long range spatial correlations (algebraic decay) in simple classical fluids 
\emph{at equilibrium }occur only near a thermodynamic critical point, i.e.
for finely tuned values of the thermodynamic parameters. On the other hand,
such long range correlations appear generically for a wide class of \emph{%
nonequilibrium} states \cite{review}. The predictions of this phenomenon
have been made in a number of contexts, including self-organized criticality %
\cite{Grinstein}, linear response \cite{response}, kinetic theory \cite%
{kinetic}, and stochastic hydrodynamics \cite{hydro}. The simplest and most
direct approach is that of linear response where the nonequilibrium
correlation functions are expanded about the equilibrium state to first
order in a nonequilibrium control parameter (typically a spatial gradient).
The algebraic decay is then seen to result from spontaneous excitations of
hydrodynamic modes induced by the coupling of the control parameter to an
associated flux. A more intuitive analysis of this effect follows from an
extension of fluctuating hydrodynamics to nonequilibrium states. To linear
order in the control parameter the same algebraic decay is found, as
expected. \ Such theoretical studies for the density autocorrelation
function in a fluid subject to a temperature gradient have received detailed
experimental confirmation in recent years \cite{Sengers}. The shortest
length scale is set by the intermolecular force range, while the
experimental verification is on macroscopic system size scales. Since the
decay is algebraic there would appear to be no other length scale involved.
However, we argue here that there is an additional macroscopic scale set by
the parameters of the\ nonequilibrium state such that the true asymptotic
decay is faster than that predicted by simple perturbative studies near
equilibrium. The analysis here is limited to a single nonequilibrium state,
that of uniform shear flow, but the qualitative features are expected to
extend to other nonequilibrium states as well.

Uniform shear flow (USF) is characterized by a constant average density and
temperature, and an average velocity flow field given by $\overrightarrow{v}%
\left( \overrightarrow{r}\right) =\overleftrightarrow{a}\cdot 
\overrightarrow{r}$ where the shear rate tensor is traceless ( $%
\overleftrightarrow{a}=ay\widehat{x}$ in a Cartesian frame of reference).
This nonequilibrium state has a single scalar control parameter $a$ and has
been the subject of numerous theoretical investigations aimed at
understanding transport and fluctuations in a model nonequilibrium state %
\cite{Machta,Tremblay,Garcia-Colin,Lutsko,Lee}. All of these share in common
the assumption that at large length and time scales, the dynamics of
fluctuations in a simple fluid are dominated by the contribution of the
hydrodynamic modes which decay much more slowly than do the neglected,
kinetic modes. The result is that the decay of thermal fluctuations in the
hydrodynamic fields, i.e. the density, momentum and energy fields, is
governed at large length and time scales by equations formally identical to
the phenomenological Navier-Stokes equations. Here, ''large length and time
scales'' means scales large compared to the mean free path and mean free
time which is the usual domain of validity of hydrodynamics. Correlations
between the values of thermally generated fluctuations in the fields at two
different space-time points can be modeled by supplementing these equations
with random forces which represent the interaction of this restricted set of
variables with the neglected degrees of freedom to give a Langevin model. In
equilibrium, the result of the Navier-Stokes-Langevin model is that the
equal-time correlation functions for two hydrodynamic fields $x\left( 
\overrightarrow{r},t\right) $ and $y\left( \overrightarrow{r},t\right) $ 
are simply proportional to delta-functions in the spatial
arguments%
\begin{equation}
C_{xy}\left( \overrightarrow{r},\overrightarrow{r}^{\prime }\right) \equiv
\left\langle \delta x\left( \overrightarrow{r},t\right) \delta y\left( 
\overrightarrow{r}^{\prime },t\right) \right\rangle \rightarrow A\delta
\left( \overrightarrow{r}-\overrightarrow{r}^{\prime }\right)   \label{0.1}
\end{equation}%
where $\delta x=x-\left\langle x\right\rangle $, and the amplitude $A$ is a
corresponding thermodynamic response function. This result simply confirms
that fluctuations at different points in space are uncorrelated when
speaking of hydrodynamic length scales (i.e., neglecting correlations on the
scale of the force range). In contrast, these correlation functions for the
nonequilibrium state have a new long range component, which to first order
in the shear rate is of the form%
\begin{equation}
C_{xy}\left( \overrightarrow{r},\overrightarrow{r}^{\prime }\right)
\rightarrow A\delta \left( \overrightarrow{r}-\overrightarrow{r}^{\prime
}\right) +aB\frac{1}{\left| \overrightarrow{r}-\overrightarrow{r}^{\prime
}\right| }  \label{0.2}
\end{equation}%
The amplitude $B$ is again a thermodynamic response function. The physical
difference between equilibrium and nonequilibrium is due to the way in which
small hydrodynamic fluctuations decay. At equilibrium this occurs locally
due to viscous and thermal damping, whereas in shear flow it is convected as
well and spreads out over a length scale that varies as the speed of
convection times the time scale for viscous and thermal damping \cite%
{Machta,Tremblay}.

With the exception of reference \cite{Lutsko}, the form (\ref{0.2}) was
generally obtained using a perturbative treatment in Fourier representation
assuming that the shear rate $a$ (with dimensions of frequency) is smaller
than all other hydrodynamic frequencies, $ck$ and $\lambda _{0}k^{2}$,
where $c$ is a propagation velocity, $\lambda _{0}$ is a transport
coefficient, and $k$ is the wavevector. For fixed shear rate, this therefore
sets an upper bound on the range of separations in real space for which the
results apply. In this context, the true asymptotic behavior of the
correlation functions remains unclear. The analysis of reference \cite%
{Lutsko} has the potential to resolve this question since it is
non-perturbative and retains the two dominant effects of large shear rate:
secular effects $\approx at$ associated with convection, and shear rates
comparable to the hydrodynamic damping $a\approx \lambda _{0}k^{2}$. The
resulting correlation functions are found to have the form%
\begin{equation}
C_{xy}\left( \overrightarrow{r},\overrightarrow{r}^{\prime }\right)
\rightarrow A\delta \left( \overrightarrow{r}-\overrightarrow{r}^{\prime
}\right) +aB\frac{1}{\left| \overrightarrow{r}-\overrightarrow{r}^{\prime
}\right| }F\left( \left( \overrightarrow{r}-\overrightarrow{r}^{\prime
}\right) /l\right) .  \label{0.3}
\end{equation}%
The function $F\left( \overrightarrow{r}/l\right) $ depends on a new
nonequilibrium correlation length, $l=\sqrt{\lambda _{0}/a}$ . For $r<<l$
the result (\ref{0.2}) is recovered. However, the asymptotic form for $r>>l$
was not explored in any detail in reference \cite{Lutsko}.

Here, we attempt to clarify the asymptotic behavior of the static
correlation functions in a sheared fluid in two ways. First, it is shown
that the continuity equation and stationarity place exact constraints on the
decay of the density autocorrelation function such that it must be faster
than $r^{-1}$ for large $r$. This result is independent of any model for
evaluating the correlation function. Next, we reconsider $F\left( 
\overrightarrow{r}/l\right) $ in (\ref{0.3}) from the results of reference %
\cite{Lutsko} and show the actual asymptotic behavior is $r^{-11/3}$. The
crossover between the $r^{-1}$ behavior at short length-scales and the
stronger $r^{-11/3}$ decay at large separations is illustrated by numerical
evaluation of the general result.

\section{Exact bounds on the rate of decay}

Consider $N$ atoms with positions and momenta denoted $\overrightarrow{q}%
_{i} $ and $\overrightarrow{p}_{i}$ respectively and denoted collectively as 
$\Gamma =\left\{ \overrightarrow{q}_{i},\overrightarrow{p}_{i}\right\}
_{i=1}^{N}$. The atoms interact via a central two-body potential $\phi
\left( r\right) $ and are confined to a volume $V$ such that the average
density is $n$. The potential is assumed to be repulsive at short distances
and to diverge as $r\rightarrow 0$ and to have a finite force range. Uniform
shear flow results from the application of Lees-Edwards boundary conditions %
\cite{Lees} consisting of periodic boundaries in all directions except that
of the gradient (here, the $y$-direction). If a particle exits the volume in
the positive $y$-direction (say, at $y=L/2$), it is re-entered at the
opposite side of the volume ( $y=-L/2$) with its velocity in the direction
of flow shifted as $v_{x}\rightarrow v_{x}-aL$. Taken together, these
constitute periodic boundaries in the local rest frame. The temperature will
generally increase due to viscous heating but we will follow standard
practice and assume the presence of a thermostat which counteracts the
viscous heating so that the temperature is also constant in time leading to
a steady-state \cite{Hoover}. The dynamics thus described possess several
important symmetries that will be used below. First, they are invariant with
respect to parity of the positions and momenta so that the equations of
motion and boundary conditions are the same if written in terms of the
variables $\widetilde{\overrightarrow{q}}_{i}=-\overrightarrow{q}_{i}$ and $%
\widetilde{\overrightarrow{p}}_{i}=-\overrightarrow{p}_{i}$. Second, they
are invariant under mirror reflection about the $z$-axis (but not the $x$ or 
$y$ axes since the boundary condition couples the $x$-velocity and the $y$%
-coordinate). Third, since USF is a steady state, statistical properties are
invariant with respect to a change in the origin from which time is measured
(time translational invariance). Fourth, USF is translationally invariant in
the local rest-frame as well as in a mixed frame consisting of the
laboratory positions and the velocities measured relative to the flow \cite%
{Lutsko}.

The microscopic local density and momentum fields are defined respectively as%
\begin{eqnarray}
\psi _{n}\left( \overrightarrow{r};\Gamma (t;\Gamma _{0})\right) 
&=&\sum_{i=1}^{N}\delta \left( \overrightarrow{r}-\overrightarrow{q}%
_{i}(t)\right)   \nonumber \\
\overrightarrow{\psi }_{p}\left( \overrightarrow{r};\Gamma (t;\Gamma
_{0})\right)  &=&\sum_{i=1}^{N}\overrightarrow{p}_{i}^{\prime }(t)\delta
\left( \overrightarrow{r}-\overrightarrow{q}_{i}(t)\right)   \label{1}
\end{eqnarray}%
where $\overrightarrow{p}_{i}^{\prime }=\overrightarrow{p}_{i}-m_{i}%
\overleftrightarrow{a}\cdot \overrightarrow{q}_{i}$ is the momentum defined
relative to the local flow field and $\Gamma (t;\Gamma _{0})$ is the point
in phase space the system would reach after evolving from the initial point $%
\Gamma _{0}$ for a time $t$. Note that these fields are related by the
microscopic continuity equation%
\begin{eqnarray}
\frac{d}{dt}\psi _{n}\left( \overrightarrow{r};\Gamma (t;\Gamma _{0})\right)
&=&\sum_{i=1}^{N}\overrightarrow{p}_{i}(t)\cdot \frac{d}{d\overrightarrow{q}%
(t)}\delta \left( \overrightarrow{r}-\overrightarrow{q}_{i}(t)\right)  
\nonumber \\
&=&-\overrightarrow{\nabla }\cdot \left( \overrightarrow{\psi }_{p}\left( 
\overrightarrow{r};\Gamma (t;\Gamma _{0})\right) +\overleftrightarrow{a}%
\cdot \overrightarrow{r}\psi _{n}\left( \overrightarrow{r};\Gamma (t;\Gamma
_{0})\right) \right) .  \label{2}
\end{eqnarray}%
The correlations functions are defined as%
\begin{equation}
C_{ab}\left( \overrightarrow{r},\overrightarrow{r}^{\prime };t,t^{\prime
}\right) \equiv \int d\Gamma _{0}\;\rho \left( \Gamma _{0}\right) \delta
\psi _{a}\left( \overrightarrow{r};\Gamma (t;\Gamma _{0})\right) \delta \psi
_{b}\left( \overrightarrow{r}^{\prime };\Gamma (t^{\prime };\Gamma
_{0})\right)   \label{3}
\end{equation}%
where subscripts $a$ and $b$ label the specific field considered, $\rho
\left( \Gamma _{0}\right) $ is the distribution of phase variables at
initial time $t=0$, and $\delta \psi _{a}=\psi _{a}-\left\langle \psi
_{a}\right\rangle $. For a steady-state the time dependence occurs through $%
t-t^{\prime }$ due to time translational invariance. Because of the modified
spatial translational invariance, it is also possible to show that the
correlation functions depend only on the relative separation. Consequently
the correlation functions can be written%
\[
C_{ab}\left( \overrightarrow{r},\overrightarrow{r}^{\prime };t,t^{\prime
}\right) =C_{ab}\left( \overrightarrow{r}-\overrightarrow{r}^{\prime
};t-t^{\prime }\right) 
\]

Choosing $\overrightarrow{r}^{\prime }=\overrightarrow{0}$ and $t^{\prime }=t
$, Eq.(\ref{3}) gives the relationship%
\begin{equation}
\frac{d}{dt}C_{nn}\left( \overrightarrow{r}\right) =0=-\overrightarrow{%
\nabla }\cdot \left( \overrightarrow{C}_{pn}\left( \overrightarrow{r}\right)
-\overrightarrow{C}_{np}\left( \overrightarrow{r}\right) +%
\overleftrightarrow{a}\cdot \overrightarrow{r}C_{nn}\left( \overrightarrow{r}%
\right) \right)   \label{4}
\end{equation}%
where for notational simplicity $C_{nn}\left( \overrightarrow{r}\right)
\equiv C_{nn}\left( \overrightarrow{r};0\right) $. This result has been
derived previously in a different context \cite{Lutsko}. It simplifies
further using the translational, parity, and reflection invariance noted
above to give the final form of interest here 
\begin{equation}
0=-\overrightarrow{\nabla }\cdot \left( 2\overrightarrow{C}_{pn}\left( 
\overrightarrow{r}\right) +\overleftrightarrow{a}\cdot \overrightarrow{r}%
C_{nn}\left( \overrightarrow{r}\right) \right) .  \label{4.3}
\end{equation}%
This is an exact result that follows directly from stationarity and
conservation of mass. Integrating (\ref{4.3}) over a spherical volume
bounded by shells at $r=0^{+}$ and $r=R$, and making use of Gauss's theorem
gives the relationship of $\overrightarrow{C}_{pn}\left( \overrightarrow{r}%
\right) $ to $C_{nn}\left( \overrightarrow{r}\right) $ 
\begin{equation}
-2\int \overrightarrow{C}_{pn}\left( R\widehat{r}\right) \cdot \widehat{r}d%
\widehat{r}=\int \widehat{r}\cdot \overleftrightarrow{a}\cdot 
\overrightarrow{r}C_{nn}\left( R\widehat{r}\right) d\widehat{r}  \label{5}
\end{equation}%
The notation $d\widehat{r}$ indicates a surface integral over the unit
sphere and use has been made of the fact that the correlation functions
evaluated at the origin vanish since the potential will not allow atoms to
occupy the same spatial position (the singular contribution to $%
C_{nn}\propto \delta \left( r\right) $ is excluded from the integration
volume).

Equation (\ref{5}) is the main result of this section. To put this in
context, consider an expansion of $C_{nn}\left( \overrightarrow{r}\right) $
to first order in the shear rate with the form 
\begin{equation}
C_{nn}\left( \overrightarrow{r}\right) \rightarrow C_{nn}^{(0)}\left(
r\right) +\widehat{r}\cdot \overleftrightarrow{a}\cdot \widehat{r}%
C_{nn}^{(1)}\left( r\right) +o(a^{2}).  \label{6}
\end{equation}%
Inserting this into eq.(\ref{5}) gives%
\begin{equation}
-2\int \overrightarrow{C}_{pn}\left( R\widehat{r}\right) \cdot \widehat{r}\;d%
\widehat{r}\rightarrow a\frac{4\pi }{15}RC_{nn}^{\left( 1\right) }\left(
R\right) +o\left( a^{2}\right) ,  \label{7}
\end{equation}%
showing that if the density-momentum correlation function decays for large
separations, as it must, then the first-order correction to the
density-density correlation function must decay faster than $1/R$ . This
result is independent of the dimensionality of the system. Furthermore,
multiplying eq.(\ref{5}) by $R^{D-1}$ and integrating gives, in $D$%
-dimensions,%
\begin{eqnarray}
\int \overrightarrow{C}_{pn}\left( \overrightarrow{r}\right) \cdot \widehat{r%
}\;d^{D}r &=&\int_{\Omega }\widehat{r}\cdot \overleftrightarrow{a}\cdot 
\widehat{r}rC_{nn}\left( \overrightarrow{r}\right) \;d^{D}r  \nonumber \\
&=&a\frac{4\pi }{15}\int_{0}^{\infty }r^{D}C_{nn}^{\left( 1\right) }\left(
r\right) dr+o(a^{2}).  \label{8}
\end{eqnarray}%
The quantity on the left is 
\begin{equation}
\int \overrightarrow{C}_{pn}\left( \overrightarrow{r}\right) \cdot \widehat{r%
}\;d^{D}r=\left\langle \sum_{i<j}\overrightarrow{p}_{ij}^{\prime }\cdot 
\widehat{q}_{ij}\right\rangle  \label{9}
\end{equation}%
which vanishes in equilibrium but can be finite for USF due to
velocity correlations. This in turn implies that the first order correction
to the shear rate decays faster than $1/r^{D+1}$.

These results can be given a somewhat more general interpretation. Any
function of a vector in terms of the spherical harmonics in order to
separate the dependence of the function on the direction and magnitude of
its argument. The density-density correlation function becomes 
\begin{equation}
C_{nn}\left( \overrightarrow{r}\right) =\sum_{L=0}^{\infty
}\sum_{M=-L}^{L}C_{nn}^{LM}\left( r\right) Y_{LM}\left( \widehat{r}\right) .
\label{10}
\end{equation}%
with%
\begin{equation}
C_{nn}^{LM}\left( r\right) =\int d\widehat{r}\;Y_{LM}^{\ast }\left( \widehat{%
r}\right) C_{nn}\left( \overrightarrow{r}\right) .  \label{11}
\end{equation}%
Because of the parity symmetry of USF, it is easy to show that only
coefficients with even values of $L$ are nonzero while the inversion
symmetry about the $z$-axis implies that only even values of $M$ contribute.
Then, noting that 
\begin{equation}
\widehat{r}_{x}\widehat{r}_{y}=-i\sqrt{\frac{2\pi }{15}}\left( Y_{22}\left( 
\widehat{r}\right) -Y_{22}^{\ast }\left( \widehat{r}\right) \right) 
\label{12}
\end{equation}%
Eq.(\ref{5}) becomes%
\begin{equation}
\int \overrightarrow{C}_{pn}\left( R\widehat{r}\right) \cdot \widehat{r}d%
\widehat{r}=-2a\sqrt{\frac{2\pi }{15}}R\func{Im}C_{nn}^{22}\left( r\right) .
\label{13}
\end{equation}%
The conclusions drawn about the first-order correction $C_{nn}^{(1)}\left(
r\right) $ are seen to be exact statements also about $\func{Im}%
C_{nn}^{22}\left( r\right) ,$ valid to \emph{all} orders in the shear rate:
namely, that $\func{Im}C_{nn}^{22}\left( r\right) $ must decay faster than $%
1/r$ in any number of dimensions and faster than $1/r^{D+1}$ in $D$ -
dimensions if the spatial integral of the radial part of the
density-momentum correlation function is finite. This corresponds precisely
to the quantity for which the first-order results (\ref{0.2}) predicted a $%
1/r$ decay. Consequently, that result cannot be correct for sufficiently
large $r$.

\section{Approximate evaluation of $C_{nn}\left( \protect\overrightarrow{r}%
\right) $}

The algebraic decays of both static and dynamic correlation functions
observed in simple fluids can be derived by means of fluctuating
hydrodynamics. In this model, the exact conservation laws for the local
density, momentum and energy density fields, $\rho ,\overrightarrow{p}$ and $%
u$ respectively, 
\begin{eqnarray}
\frac{\partial }{\partial t}\rho +\overrightarrow{\nabla }\cdot 
\overrightarrow{p} &=&0  \nonumber \\
\frac{\partial }{\partial t}\overrightarrow{p}+\overrightarrow{\nabla }\cdot 
\overleftrightarrow{P} &=&0  \nonumber \\
\frac{\partial }{\partial t}u+\overrightarrow{\nabla }\cdot \overrightarrow{q%
} &=&0  \label{13a}
\end{eqnarray}%
are approximated by taking the pressure tensor, $\overleftrightarrow{P},$
and heat flux vector, $\overrightarrow{q}$, to be a sum of two terms: the
usual Navier-Stokes functionals of the local fields, and a random component
which is delta-function correlated in space and time. The amplitudes of
these correlations are related to the forms of the deterministic parts of
the fluxes \cite{fluctuation}. Rewriting these in terms of deviations from
the macroscopic state then gives a description of fluctuations about this
state. The details for USF have been discussed in detail elsewhere \cite%
{Lutsko} and only the results are quoted for the purposes here. Defining the
Fourier transform of the density-density correlation function as%
\begin{equation}
\widetilde{C}_{nn}\left( \overrightarrow{k}\right) =\int d\overrightarrow{r}%
\;\exp \left( i\overrightarrow{k}\cdot \overrightarrow{r}\right)
C_{nn}\left( \overrightarrow{r}\right)   \label{14}
\end{equation}%
the result obtained for it is 
\begin{equation}
\widetilde{C}_{nn}\left( \overrightarrow{k};a\right) =k_{B}T_{0}\rho
_{0}^{2}\chi _{T}\left( 1+\gamma ^{-1}\widetilde{\Delta }_{nn}\left( 
\overrightarrow{k}l\right) \right)   \label{16}
\end{equation}%
\begin{equation}
\widetilde{\Delta }_{nn}\left( \overrightarrow{k}\right) =\int_{0}^{\infty
}ds\frac{kk_{x}k_{y}\left( -s\right) }{k^{3}(-s)}\exp \left(
-\int_{0}^{s}ds^{\prime }\;k^{2}\left( -s^{\prime }\right) \right) .
\label{16.1}
\end{equation}%
Here $\chi _{T}$ is the isothermal compressibility, $c_{0}$ is the
equilibrium speed of sound, $\lambda _{0}$ is the equilibrium sound-damping
constant, $\gamma =c_{p}/c_{V}$ is the ratio of specific heats at constant
pressure and volume, and $\overrightarrow{k}(t)=\left(
k_{x},k_{y}-tk_{x},k_{z}\right) $. The characteristic length scale is $l=%
\sqrt{\lambda _{0}/a}$ . The first term in eq.(\ref{16}) is the equilibrium
contribution which, in the small wavevector approximation used here, is a
constant. The second term represents the nonequilibrium correction which is
derived assuming that the shear rate and rate of dissipation are
significantly less than the sound frequency, $a,\lambda _{0}k^{2}\ll c_{0}k$
where $c_{0}$ is the speed of sound. However, no restriction on the value of 
$a/\lambda _{0}k^{2}$ is imposed. The inverse transform of eq. (\ref{16}) is
the real space result quoted in (\ref{0.3}) above.

In order to evaluate the behavior in real space, it is useful to introduce
the expansion of $\widetilde{C}_{nn}\left( \overrightarrow{k}\right) $ in
spherical harmonics 
\begin{equation}
\widetilde{C}_{nn}\left( \overrightarrow{k}\right) =\sum_{L=0}^{\infty
}\sum_{M=-L}^{L}\overline{C}_{nn}^{LM}\left( k\right) Y_{LM}\left( \widehat{k%
}\right)   \label{17}
\end{equation}%
and to recall the relation between the coefficients (\ref{11}) to those in (%
\ref{17})%
\begin{equation}
C_{nn}^{LM}\left( r\right) =\frac{1}{2\pi ^{2}i^{L}}\int_{0}^{\infty
}k^{2}dk\;j_{L}(kr)\overline{C}_{nn}^{LM}\left( k\right) .  \label{18}
\end{equation}%
This gives directly%
\begin{equation}
C_{nn}^{LM}\left( r\right) =\left( k_{B}T_{0}\rho _{0}^{2}\chi _{T}\right) 
\sqrt{4\pi }l^{-3}\delta \left( \overrightarrow{r}/l\right) \delta
_{L0}\delta _{M0}+\left( k_{B}T_{0}\rho _{0}^{2}\chi _{T}\gamma ^{-1}\right) 
\frac{1}{2\pi ^{2}i^{L}}l^{-3}\Delta _{nn}^{LM}\left( r/l\right)   \label{19}
\end{equation}%
with%
\begin{eqnarray}
\Delta _{nn}^{LM}\left( \rho =r/l\right)  &=&\int d\overrightarrow{k}%
\;j_{L}\left( k\rho \right) Y_{LM}^{\ast }\left( \widehat{k}\right)
\int_{0}^{\infty }ds\frac{kk_{x}k_{y}\left( -s\right) }{k^{\prime 3}(-s)}%
\exp \left( -\int_{0}^{s}ds^{\prime }\;k^{2}\left( -s^{\prime }\right)
\right)   \nonumber \\
&=&\left( \frac{1}{\rho }\right) \int d\overrightarrow{k}\;j_{L}\left(
k\right) Y_{LM}^{\ast }\left( \widehat{k}\right)   \nonumber \\
&&\times \int_{0}^{\infty }ds\frac{kk_{x}k_{y}\left( -s\rho ^{2}\right) }{%
k^{3}(-s\rho ^{2})}\exp \left( -\int_{0}^{s}ds^{\prime }\;k^{2}\left(
-s^{\prime }\rho ^{2}\right) \right) .  \label{20}
\end{eqnarray}%
The small shear rate limit (\ref{0.2}) is obtained by noting that $l^{2}$ is
inversely proportional to the shear rate and expanding to leading order in $a
$%
\begin{eqnarray}
l^{-3}\Delta _{nn}^{LM}\left( r/l\right)  &\rightarrow &l^{-3}\left( \frac{l%
}{r}\right) \left[ -\int d\overrightarrow{k}\;j_{L}\left( k\right)
Y_{LM}^{\ast }\left( \widehat{k}\right) \frac{k_{x}k_{y}}{k^{4}}\right]  
\nonumber \\
&\rightarrow &\frac{a}{\lambda _{0}r}\left[ i\delta _{L2}\left( \delta
_{M2}-\delta _{M-2}\right) \frac{\pi }{4}\sqrt{\frac{2\pi }{15}}\right] 
\label{20b}
\end{eqnarray}%
The context here shows that this result applys only in the limit $%
r/l\rightarrow 0$ which is to say that, for fixed shear rate, this is a 
\emph{small} $r$ result and does not represent the true long-range behavior
of the correlation function.

The actual asymptotic behavior for large $\rho =r/l$ is obtained in the
Appendix where it is found that 
\begin{eqnarray}
\Delta _{nn}^{00}\left( \rho \right) &\rightarrow &\left( \frac{2}{9}\right)
^{2/3}5\pi \Gamma \left( \frac{5}{6}\right) \rho ^{-11/3}+O\left( \rho
^{-13/3}\right)  \nonumber \\
&\approx &6.\,\allowbreak 5\text{ }\rho ^{-11/3}.  \label{21}
\end{eqnarray}
All other components decay even more rapidly. In particular the true
asymptotic behavior of $\func{Im}\Delta _{nn}^{22}\left( x\right) $ is found
to be%
\begin{eqnarray}
\func{Im}\Delta _{nn}^{22}\left( \rho \right) &\rightarrow &\frac{935}{756}%
6^{1/6}\sqrt{5}\pi \Gamma \left( \frac{5}{6}\right) \rho ^{-17/3}+O\left(
\rho ^{-19/3}\right)  \nonumber \\
&\approx &13.\,\allowbreak 2\text{ }\rho ^{-17/3}.  \label{21a}
\end{eqnarray}
This is consistent with the exact result (\ref{13}) that this component must
decay more slowly than $r^{-4}$ in three dimensions.

\begin{figure*}
\includegraphics[angle=-90,scale=0.5]{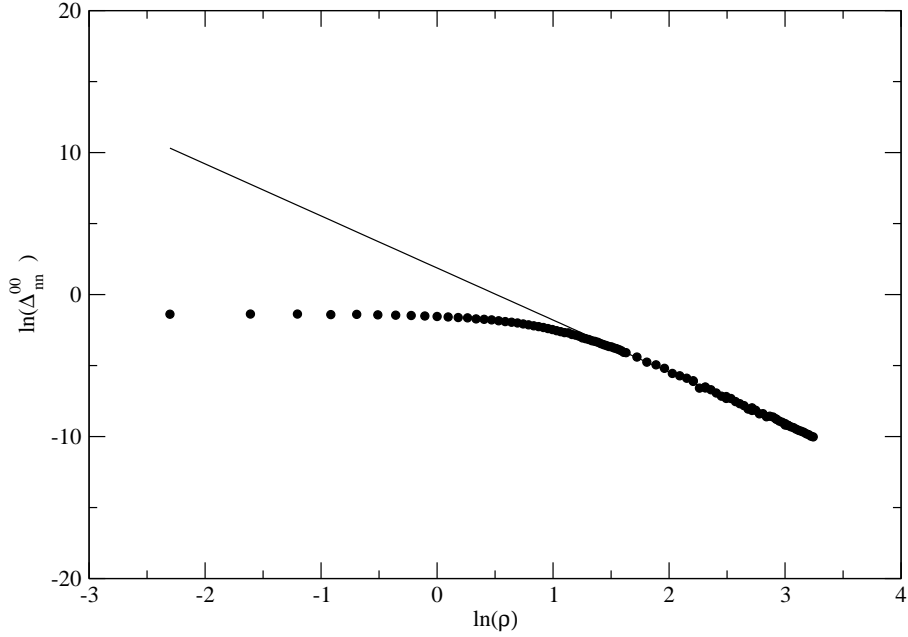}
\caption{\label{fig1} $\ln \left(\Delta _{nn}^{00}\right)$ as a function of $\ln (r/l)$ as determined by
the numerical calculation (circles) and the asymptotic result given in eq.(\ref{21}%
) (line).}
\end{figure*}

In order to probe the asymptotic behavior in more detail, we have performed
a numerical evaluation of eq.(\ref{20}) by means of multi-dimensional Monte
Carlo integration using the VEGAS algorithm\cite{Vegas,Press,GSL}. Rather
than directly evaluating eq.(\ref{20}), it was found to be more efficient to
separate out the short-ranged $1/r$ behavior by rewriting this as%
\begin{eqnarray}
\Delta _{nn}^{LM}\left( \rho \right)  &=&-i\delta _{2l}\left( \delta
_{m2}-\delta _{m-2}\right) \sqrt{\frac{2\pi }{15}}\frac{\pi }{4\rho }+\int d%
\overrightarrow{k}\;j_{L}\left( k\rho \right) Y_{LM}^{\ast }\left( \widehat{k%
}\right)   \nonumber \\
&&\times \int_{0}^{\infty }ds\left( \frac{kk_{x}k_{y}\left( -s\right) }{%
k^{3}(-s)}-\frac{k^{2}\left( -s\right) k_{x}k_{y}}{k^{4}}\right) \exp \left(
-\int_{0}^{s}ds^{\prime }\;k^{2}\left( -s^{\prime }\right) \right) .
\label{22}
\end{eqnarray}%
The number of samples used in the performing the integrals was adjusted so
that the internal estimate of the error in the evaluations was always less
than 5\% of the calculated values. For small $\rho $, the errors were
substantially less while the limit was occasionally reached as $\rho $ was
increased. Figure 1 shows the spherically averaged value, $\Delta
_{nn}^{00}\left( \rho \right) $, as a function of $\rho $ together with the
asymptotic power law of $-11/3$ and the two are seen to be consistent.
Figure 2 shows the numerical calculation $\Delta _{nn}^{22}\left( \rho
\right) $ in comparison with the small- and large-$r$ limits, eqs.(\ref{20b}%
) and (\ref{21a}) respectively. Again, crossover between the limiting forms
is clearly identified.

\begin{figure*}
\includegraphics[angle=-90,scale=0.5]{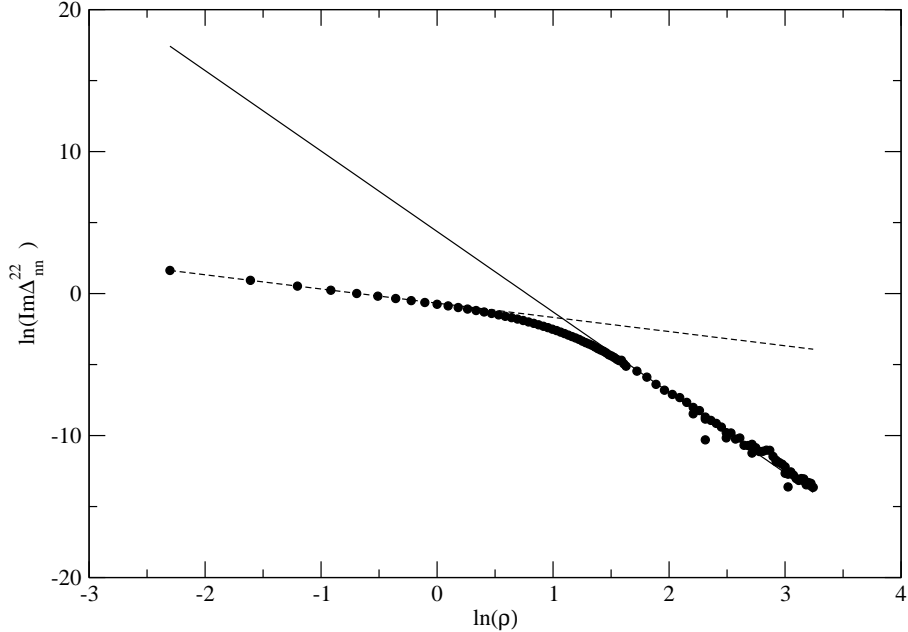}
\caption{\label{fig2} $\ln \left(Im \Delta _{nn}^{22}\right)$ as a function of $\ln (r/l)$ as determined by
the numerical calculation (circles) and the asymptotic results given in eq.(\ref{21a}%
) (full line) and eq.(\ref{20b}) (dotted line).}
\end{figure*}

\section{Discussion}

The prediction of $1/r$ decays in the density autocorrelation function in
shear flow is in violation of an exact bound coming from elementary
considerations of statistical mechanics and the properties of the \ USF
steady state. Detailed analysis of a less restrictive solution of the
Navier-Stokes-Langevin model confirms that the $1/r$ behavior is actually
valid only for $r<\sqrt{\lambda _{0}/a}$ and that this crosses over to a
stronger power law decay at large distances. It turns out for this model
that the density-energy correlation function as well as energy and
longitudinal-velocity autocorrelation functions share the same spatial
dependence given by (\ref{20b}). Therefore the calculations given here apply
to them as well. Interestingly, the most long-ranged correlation function,
based on calculations analogous to that illustrated in the appendix, is for
one of the transverse-velocity autocorrelations which, in the notation of
ref \cite{Lutsko}, has the form%
\begin{equation}
C_{44}(\overrightarrow{r})=k_{B}T_{0}\left[ 1+\Delta _{44}(\overrightarrow{r}%
/l^{\prime })\right]  \label{23}
\end{equation}%
with $l^{\prime }=\sqrt{2\nu _{0}/a}$ where $\nu _{0}$ is the shear
viscosity and 
\begin{equation}
\Delta _{44}^{00}\left( \rho \right) \rightarrow \frac{1}{5}6^{2/3}\pi
\Gamma \left( \frac{5}{6}\right) \rho ^{-5/3}+O(\rho ^{-7/3}).  \label{24}
\end{equation}

The analysis here has implicitly assumed stability of the USF state. In fact
USF is unstable to sufficiently long wavelength perturbations \cite{Lee,Lee2}%
. The critical wavelength for stability scales approximately as $v_{T}/a$
for small $a$ where $v_{T}$ is the thermal velocity. Therefore in order to
see the crossover phenomenon discussed here there must be conditions such
that $l<<v_{T}/a$. This requires $a\lambda _{0}/v_{T}^{2}<<1$ which can be
accomplished by small shear rates and high temperatures. The predicted
behavior should be accessible via molecular dynamics simulation for a
sufficiently large system.\ 

The qualitative feature of a nonequilibrium length scale should be more
general than the special case of USF, and applicable to other nonequilibrium
states as well. For example, a steady state with uniform temperature
gradient has a characteristic frequency $v_{T}\nabla \ln T$. Setting this
equal to a hydrodynamic damping gives the length scale $l=\sqrt{\lambda
_{0}/\left( v_{T}\nabla \ln T\right) }$. It is expected that the asymptotic
decay for $r>>l$ will be different from that of perturbative mode coupling
theory currently in the literature for reasons similar to those given here
for USF.

\section{Acknowledgments}

J. F. Lutsko acknowledges support from the Universit\'{e} Libre de
Bruxelles. The research of J. Dufty was supported in part by U. S.
Department of Energy grant DE-FG03-98DP00218.

\appendix 

\section{Asymptotic behaviour of the correlation functions}

We begin by writing%
\begin{equation}
\widetilde{\Delta }_{nn}\left( \overrightarrow{k}\right) =\int_{0}^{\infty
}dt\;\frac{\widehat{k}_{x}\widehat{k}_{y}+t\widehat{k}_{x}^{2}}{\left( 1+2%
\widehat{k}_{x}\widehat{k}_{y}t+\widehat{k}_{x}^{2}t^{2}\right) ^{3/2}}\exp
\left( -k^{2}\beta (t)\right)  \label{a1}
\end{equation}%
where $\beta (t)=t+\widehat{k}_{x}\widehat{k}_{y}t^{2}+\frac{1}{3}\widehat{k}%
_{x}^{2}t^{3}$ is independent of the magnitude of the wavevector. The
coefficients of the expansion in spherical harmonics in real space are then%
\begin{eqnarray}
\Delta _{nn}^{LM}\left( \rho \right) &=&\int d\widehat{k}\;Y_{LM}^{\ast
}\left( \widehat{k}\right) \int_{0}^{\infty }dkk^{2}j_{l}(k\rho
)\int_{0}^{\infty }dt\;\frac{\widehat{k}_{x}\widehat{k}_{y}+t\widehat{k}%
_{x}^{2}}{\left( 1+2\widehat{k}_{x}\widehat{k}_{y}t+\widehat{k}%
_{x}^{2}t^{2}\right) ^{3/2}}\exp \left( -k^{2}\beta (t)\right)  \nonumber \\
&=&\sqrt{\pi }\frac{\Gamma \left( \frac{L+3}{2}\right) }{2^{L+2}\Gamma
\left( \frac{2L+3}{2}\right) }\rho ^{-3}\int d\widehat{k}\;Y_{LM}^{\ast
}\left( \widehat{k}\right)  \nonumber \\
&&\times \int_{0}^{\infty }dt\;\frac{\widehat{k}_{x}\widehat{k}_{y}+t%
\widehat{k}_{x}^{2}}{\left( 1+2\widehat{k}_{x}\widehat{k}_{y}t+\widehat{k}%
_{x}^{2}t^{2}\right) ^{3/2}}\left( \frac{\rho ^{2}}{\beta (t)}\right) ^{%
\frac{L+3}{2}}M\left( \frac{L+3}{2},\frac{2L+3}{2},-\frac{\rho ^{2}}{4\beta
(t)}\right)  \label{a2}
\end{eqnarray}%
where $M(a,b,z)$ is the confluent hypergeometric function. Now, changing
variables to $t=\rho ^{2/3}\left( \widehat{k}_{x}^{2}\right) ^{-1/3}y^{-1}$
gives%
\begin{eqnarray}
\Delta _{nn}^{LM}\left( \rho \right) &=&\sqrt{\pi }\frac{\Gamma \left( \frac{%
L+3}{2}\right) }{2^{L+2}\Gamma \left( \frac{2L+3}{2}\right) }r^{-11/3}\int d%
\widehat{k}\;Y_{LM}^{\ast }\left( \widehat{k}\right) \left( \widehat{k}%
_{x}^{2}\right) ^{-1/6}  \nonumber \\
&&\times \int_{0}^{\infty }dy\;\frac{1+\widehat{k}_{x}\widehat{k}_{y}\left( 
\widehat{k}_{x}^{2}\right) ^{-2/3}y\rho ^{-2/3}}{\left( 1+2\widehat{k}_{x}%
\widehat{k}_{y}\left( \widehat{k}_{x}^{2}\right) ^{-2/3}y\rho ^{-2/3}+\left( 
\widehat{k}_{x}^{2}\right) ^{-1/3}y^{2}\rho ^{-4/3}\right) ^{3/2}}  \nonumber
\\
&&\times \left( \frac{y^{3}}{\gamma (y)}\right) ^{\frac{L+3}{2}}M\left( 
\frac{L+3}{2},\frac{2L+3}{2},-\frac{y^{3}}{4\gamma (y)}\right)  \label{a3}
\end{eqnarray}%
with%
\begin{equation}
\gamma (y)=\frac{1}{3}+\widehat{k}_{x}\widehat{k}_{y}\left( \widehat{k}%
_{x}^{2}\right) ^{-2/3}y\rho ^{-2/3}+\left( \widehat{k}_{x}^{2}\right)
^{-1/3}y^{2}\rho ^{-4/3}.  \label{a4}
\end{equation}%
For the $00$ component, we use $M\left( a,a,z\right) =\exp (z)$ to get%
\begin{eqnarray}
\Delta _{ab}^{00}\left( r\right) &=&\frac{1}{2^{3}}\rho ^{-11/3}\int d%
\widehat{k}\;\left( \widehat{k}_{x}^{2}\right) ^{-1/6}\int_{0}^{\infty
}dy\;\left( 3y^{3}\right) ^{\frac{3}{2}}\exp \left( -\frac{3y^{3}}{4}\right)
+O\left( \rho ^{-13/3}\right) \\
&=&2^{\frac{2}{3}}3^{-\frac{4}{3}}5\pi \Gamma \left( \frac{5}{6}\right) \rho
^{-11/3}+O\left( \rho ^{-13/3}\right) .  \nonumber
\end{eqnarray}

For the $22$ component, it is more convienient to go back to the beginning
and to integrate by parts%
\begin{eqnarray}
\Delta _{nn}^{LM}\left( \rho \right) &=&\int d\widehat{k}\;Y_{LM}^{\ast
}\left( \widehat{k}\right) \int_{0}^{\infty }k^{2}dkj_{L}(k\rho )  \nonumber
\\
&&\times \int_{0}^{\infty }dt\;\left[ -\frac{d}{dt}\left( 1+2\widehat{k}_{x}%
\widehat{k}_{y}t+\widehat{k}_{x}^{2}t^{2}\right) ^{-1/2}\right] \exp \left(
-k^{2}\beta (t)\right)  \nonumber \\
&=&\rho ^{-3}\int d\widehat{k}\;Y_{LM}^{\ast }\left( \widehat{k}\right)
\int_{0}^{\infty }k^{2}dkj_{L}(k)-\sqrt{\pi }\frac{\Gamma \left( \frac{L+5}{2%
}\right) }{2^{L+2}\Gamma \left( L+\frac{3}{2}\right) }\rho ^{-5}\int d%
\widehat{k}\;Y_{LM}^{\ast }\left( \widehat{k}\right)  \nonumber \\
&&\int_{0}^{\infty }dt\;\left( 1+2\widehat{k}_{x}\widehat{k}_{y}t+\widehat{k}%
_{x}^{2}t^{2}\right) ^{1/2}\left( \frac{\rho ^{2}}{\beta (t)}\right) ^{\frac{%
1}{2}L+\frac{5}{2}}M\left( \frac{L+5}{2},\frac{2L+3}{2},-\frac{\rho ^{2}}{%
4\beta (t)}\right)  \label{a5}
\end{eqnarray}%
The first term vanishes for $L>1$ so%
\begin{eqnarray}
\func{Im}\Delta _{nn}^{22}\left( \rho \right) &=&-\sqrt{\pi }\frac{1}{2^{4}}%
\rho ^{-5}\int d\widehat{k}\;\func{Im}Y_{22}^{\ast }\left( \widehat{k}\right)
\nonumber \\
&&\times \int_{0}^{\infty }dt\;\left( 1+2\widehat{k}_{x}\widehat{k}_{y}t+%
\widehat{k}_{x}^{2}t^{2}\right) ^{1/2}\left( \frac{\rho ^{2}}{\beta (t)}%
\right) ^{\frac{7}{2}}\exp \left( -\frac{\rho ^{2}}{4\beta (t)}\right)
\label{a6}
\end{eqnarray}%
and making the same change of variables as above gives 
\begin{eqnarray}
\func{Im}\Delta _{nn}^{22}\left( \rho \right) &=&-\sqrt{\pi }\frac{1}{2^{4}}%
\rho ^{-11/3}\int d\widehat{k}\;\func{Im}Y_{22}^{\ast }\left( \widehat{k}%
\right) \left( \widehat{k}_{x}^{2}\right) ^{-1/6}  \nonumber \\
&&\times \int_{0}^{\infty }y^{-3}dy\;\left( 1+2\widehat{k}_{x}\widehat{k}%
_{y}\rho ^{-2/3}\left( \widehat{k}_{x}^{2}\right) ^{-2/3}y+\rho
^{-4/3}\left( \widehat{k}_{x}^{2}\right) ^{-1/3}y^{2}\right) ^{1/2} 
\nonumber \\
&&\times \left( \frac{y^{3}}{\gamma (y)}\right) ^{\frac{7}{2}}\exp \left( -%
\frac{y^{3}}{4\gamma (y)}\right)  \label{a7}
\end{eqnarray}%
Now, only the odd terms in $\widehat{k}_{x}\widehat{k}_{y}$ give
nonvanishing contributions so we expand as

\begin{eqnarray}
&&\left( 1+2\widehat{k}_{x}\widehat{k}_{y}\rho ^{-2/3}\left( \widehat{k}%
_{x}^{2}\right) ^{-2/3}y+\rho ^{-4/3}\left( \widehat{k}_{x}^{2}\right)
^{-1/3}y^{2}\right) ^{1/2}\left( \frac{y^{3}}{\gamma (y)}\right) ^{\frac{7}{2%
}}\exp \left( -\frac{y^{3}}{4\gamma (y)}\right)  \nonumber \\
&=&\frac{27}{4}\sqrt{3}\widehat{k}_{x}\widehat{k}_{y}\left( \widehat{k}%
_{x}^{2}\right) ^{-2/3}\left( 9y^{3}-38\right) y^{\frac{23}{2}}\allowbreak
\exp \left( -\frac{3}{4}y^{3}\right) \rho ^{-2/3}+\frac{27}{128}\sqrt{3}y^{%
\frac{27}{2}}\widehat{k}_{x}\widehat{k}_{y}\left( \widehat{k}_{x}^{2}\right)
^{-2}\exp \left( -\frac{3}{4}y^{3}\right) \allowbreak \allowbreak \rho ^{-2}
\nonumber \\
&&\times \left( 8\left( \widehat{k}_{x}^{2}\right) \left(
-918y^{3}+81y^{6}+2008\right) +\left( \widehat{k}_{x}\widehat{k}_{y}\right)
^{2}\left( 28\,044y^{3}-5022y^{6}+243y^{9}-40\,088\right) \allowbreak \right)
\nonumber \\
&&+O(r^{-8/3})+\text{even}  \label{a8}
\end{eqnarray}

$\allowbreak $The $y$-integral of the first term vanishes and the next term
gives%
\begin{eqnarray}
\func{Im}\Delta _{nn}^{22}\left( \rho \right) &=&\frac{935}{729}3^{\frac{2}{3%
}}\sqrt{\pi }2^{\frac{2}{3}}\Gamma \left( \frac{5}{6}\right) \rho ^{-17/3} 
\nonumber \\
&&\times \int d\widehat{k}\;\func{Im}Y_{lm}^{\ast }\left( \widehat{k}\right)
\left( \widehat{k}_{x}^{2}\right) ^{-1/6}\widehat{k}_{x}\widehat{k}%
_{y}\left( \widehat{k}_{x}^{2}\right) ^{-2}\left( 3\left( \widehat{k}%
_{x}^{2}\right) -2\left( \widehat{k}_{x}\widehat{k}_{y}\right) ^{2}\right) 
\nonumber \\
&=&\frac{935}{756}3^{\frac{2}{3}}\sqrt{30}2^{\frac{2}{3}}\pi \Gamma \left( 
\frac{5}{6}\right) \rho ^{-17/3}  \label{a9}
\end{eqnarray}

\end{document}